\documentclass[12pt,letterpaper]{article}
\usepackage{amsmath,amssymb,amsthm,subfigure} 
\usepackage{dcolumn} 
\usepackage{bm} 
\usepackage{xcolor} 
\usepackage{psfrag} 
\usepackage[dvips]{graphicx}

\begin{document}
\title{On the non-physical concavity of the quark potentials within the thick center vortex model}
\author{H.~Lookzadeh\footnote{E-mail: h.lookzadeh@yazd.ac.ir }  and M.~Hosseini\\
\begin{scriptsize}
Faculty of Physics, Yazd University, P.O. Box 89195-741, Yazd, Iran
\end{scriptsize}}
\maketitle 
\begin{abstract}
Lattice gauge theory results show the confinement for the quark potential in different Yang-Mills theories and even the G(2) gauge theory. LGT calculations show that quark potential should have the down concavity behaviour. Confinement properties can be explained using the thick center vortex model. However an upward concavity is seen in the quark potential intervals using this model. After study the reason of this concavity, it is shown the non-physical concavity can be reduced by taking an arbitrary symmetric vortex flux in the space time plane of the lattice. 
\end{abstract}

\section{Introduction }
The problem of the quantum chromodynamics (QCD) vacuum properties especially confinement is interesting in the particle physics \cite{shuryak, Greensite2011}. There is not yet an analytic proof of color confinement in any non-abelian gauge theory. Non-linear properties of the confinement make it hard to study using the usual perturbation quantum field theory such as Feynman diagrams techniques. So, non-perturbation methods are used to study this phenomenon. Lattice gauge theory is a useful method to explain such non-linear phenomenon\cite{creutz1983}. The confining phase is usually defined by the behaviour of the action of the Wilson loop. Wilson loop is simply the path in space-time traced out by a quark antiquark pair created at one point and annihilated at another point\cite{wilson1974}. In a non-confining theory, the action of such a loop is proportional to its perimeter. However, in a confining theory, the action of the loop instead of its perimeter is proportional to its area. Since the area is proportional to the separation of the quark antiquark pair, free quarks are suppressed. Mesons are allowed in such a picture, since a loop containing another loop with the opposite orientation has only a small area between the two loops. Topological properties of non abelian theories seem interesting for studying the quark confinement. Phenomenological models are used to study the confinement with different approaches. In these models, the QCD vacuum is filled with some topological configurations which are confining coloured objects. The most popular candidates among these topological fields are monopoles and vortices \cite{topology}. Other candidates include merons, calorons, etc. There are very strong correlations between these various objects, though. 
The center vortex model was initially introduced by 't Hooft \cite{Hooft1978, Hooft1979}. It is able to explain the confinement of quark pairs at the asymptotic region, but it is not able to explain the confinement at intermediate distances especially for the higher representations. The model then is modified to the thick center vortex model \cite{Faber1998}. Within this model one can obtain the Casmir scaling and N-ality behaviour of the gauge theory. This is done before for the $SU(2)$, $SU(3)$ and $SU(4)$ \cite{Deldar2001, Deldar2005, Rafibakhsh2007, Bali2000, Deldar2000}. Using another modification of the model to the domain vacuum structures is possible to describe the properties of gauge theory without non trivial center elements such as the G(2) gauge theory \cite{Lookzadeh2012}. 
  
One of the properties of the quark potentials is downward concavity. Due to this condition it is not possible to observe an upward concavity in the quark potential. However using the thick center vortex model such behaviour is observed. In this article after studying a proof for this condition the reason of presence of such non-physical properties is understood more. Due to this study considering an arbitrary symmetric vortex flux in the space time is suggested to reduce the concavity. In the next section the thick center vortex is introduced. In the section $III$ a proof for the concavity of the quark potential using the lattice gauge theory is introduced. In the section $IV$ it is shown that an arbitrary symmetric vortex flux is needed to avoid the concavity of the potential. Also in the section $IV$ a method to avoiding the quark concavity is introduced and then applied to the thick center vortex model. Elimination of the concavity is shown for the $SU(2)$ and $SU(3)$ in different representations in the Casimir region and a whole reduction is observed. The section $V$ is devoted to the Casimir scaling properties of the Casimir using this method. 

\section{Thick Center vortex Model }

\begin{figure*} [!ht]
\begin{center}
\includegraphics [scale=0.8]{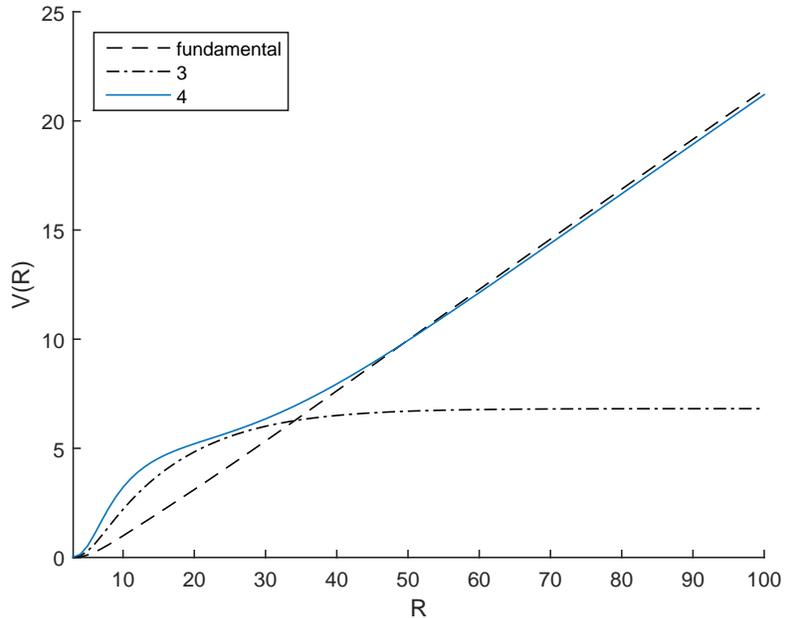}
\caption{ 
Potentials between static sources of $ SU(2)$ gauge group using thick center vortex model. At some intervals, specially for representation $4$ at the distances $10 \leqslant R  \leqslant  30$ a non physical concavity is observed.}
\label{su2tak}
\end{center}
\end{figure*}

The confinement mechanism can be related to the center elements of a gauge group. This is done through the center vortex idea. Vortices here are some 1+1 dimensional soliton like solution which is embedded in 1+3 dimension. They form closed surfaces which can be linked to Wilson loop. In the center vortex picture presence of vortices in the vacuum is due to the center elements and their fluctuation in the number of center vortices linked to the loop lead to an area law Wilson loop and a linear potential or string like behaviour. Wilson loops are gauge-invariant observable obtained from the holonomy of the gauge connection around given loops. Confinement is obtained from random fluctuations in the linking number. A vortex piercing a Wilson loop contributes to a center element Z somewhere between the group elements U of the gauge group. The Wilson loop becomes 
\begin{equation}\label{W}
W(C)=Tr[UUU...U ]\longrightarrow Tr[UU....(Z)U]
\end{equation}
Center elements commute with all elements of the group, so the location of Z in eq. (\ref{W}) can be changed by changing the place of discontinuity which lead to a vortex formation. In the $SU(2)$ group for example the string tension $\sigma$ can be obtained considering the vacuum expectation value of the Wilson loop:
\begin{equation}\label{VEV}
\langle W(C)\rangle =\prod \lbrace (1-f)+f(-1) \rbrace \langle W_{0}(C)\rangle \\
=exp[-\sigma(C) A] \langle W_{0}(C)\rangle.
\end{equation}
$f$ here is the probability of piercing a plaquette with a thin vortex some where on a Wilson loop and $W_{0}(C)$ is the Wilson loop with no linking to a vortex. $A$ is the area of the Wilson loop and is equal to $R\times T$ . $R$ is for the space side and $T$ time side of the Wilson loop. String tension can be obtained as
\begin{equation}
\sigma =\frac{-1}{A}ln(1-2f).
\end{equation}
Center vortex scenario can explain the asymptotic string behaviour in different representations of the gauge group, but cannot explain the intermediate behaviour of the quark potential. In this scenario the vortices are considered thin. However LGT results show that these vortex structures have a comparable thickness. Having an explanation for these properties a thickness must be considered for the vortices. This is done by considering the parameter $G$ instead of the center element $Z$ such that
\begin{equation}\label{G}
G(x,s)=S exp(i\alpha_{C}(x)\vec{n}.\vec{L})S^{\dagger}
\end{equation} 
$L_{i}$ ’s are the generators of the group in the representation $j$, $n$ is a unit vector, and $S$ is an element of the group $SU(N)$ in the representation $j$. $\alpha_{C}(x)$ gives the profile of the vortex, and it depends on that fraction of the vortex which is pierced by the loop. It depends on the shape of the loop $C$ and the position of the center of the vortex relative to the perimeter of the loop. 
The Wilson loop expectation value considering the thickness is obtained as 
\begin{equation}
V(R)=\sum_{x}Ln\lbrace 1-\sum_{n=1}^{N-1}f_{n}(1-Reg_{r}[\vec{\alpha}_{C}^{n}(x)]\rbrace,
\end{equation}
In which $ g_{r}[\vec{\alpha}_{C}^{n}(x)]$ is obtained by averaging over group space direction as follows 
\begin{equation}
g_{r}[\vec{\alpha}]=\frac{1}{d_r}Tr(exp[i\vec{\alpha}.\vec{H}])
\end{equation}
$H$s are the diagonal generators of the representation of the group. For the vortex flux the following ansatz is considered
\begin{equation}
\vec{\alpha}_{C}^{n}(x)=\vec{N}^n[1-tanh(ay(x)+\frac{b}{R})]
\end{equation}
$a,b$ are the parameters of the model and $y(x)$ is 
\begin{equation}
y(x)=\lbrace \begin{array}{cc} -x & \lvert R-x\lvert>|x| \\x-R & \lvert R-x\lvert\leqslant |x| \end{array},
\end{equation}
$y(x)$ is the nearest distance of the vortex center $x$, from time like side of the Wilson loop. The normalization constant $\vec{N}^n$ is obtained from the maximum flux condition, where the loop contains the vortex completely,
\begin{equation}
exp(i\vec{\alpha}.\vec{H})=z_nI
\end{equation}
$z_n$ is 
\begin{equation}
z_n=exp(\frac{2\pi in}{N})\in Z_n.
\end{equation}
$I$ is the unit element of the group. $Z_n$s are the center elements of the group.
Using this model the quark potential in different gauge group can be obtained. Here the model is applied to the simplest non abelian group $SU(2)$ and the QCD color symmetry $SU(3)$. To obtain the quark potential using this model the diagonal generators are needed and the normalization constant should be obtained. The fundamental representation generators are proportional to the Pauli matrices for the $SU(2)$ group. The free parameters of  the model are considered as $a=0.05, b=4, f=0.1$. The quark potentials behaviour using this model is shown in Figure \ref{su2tak} . For the fundamental representation and also the adjoint and also $4$ representations the potentials are obeying the $2$-ality in the final quark potential behaviour. An upward concavity is observed in the quark potential at distances 10 to 30 which is not physical. According to the quark potential condition it is not possible for the quark potential to have an upside concavity. 

\begin{figure*} [!ht]
\begin{center}
\includegraphics [scale=0.8] {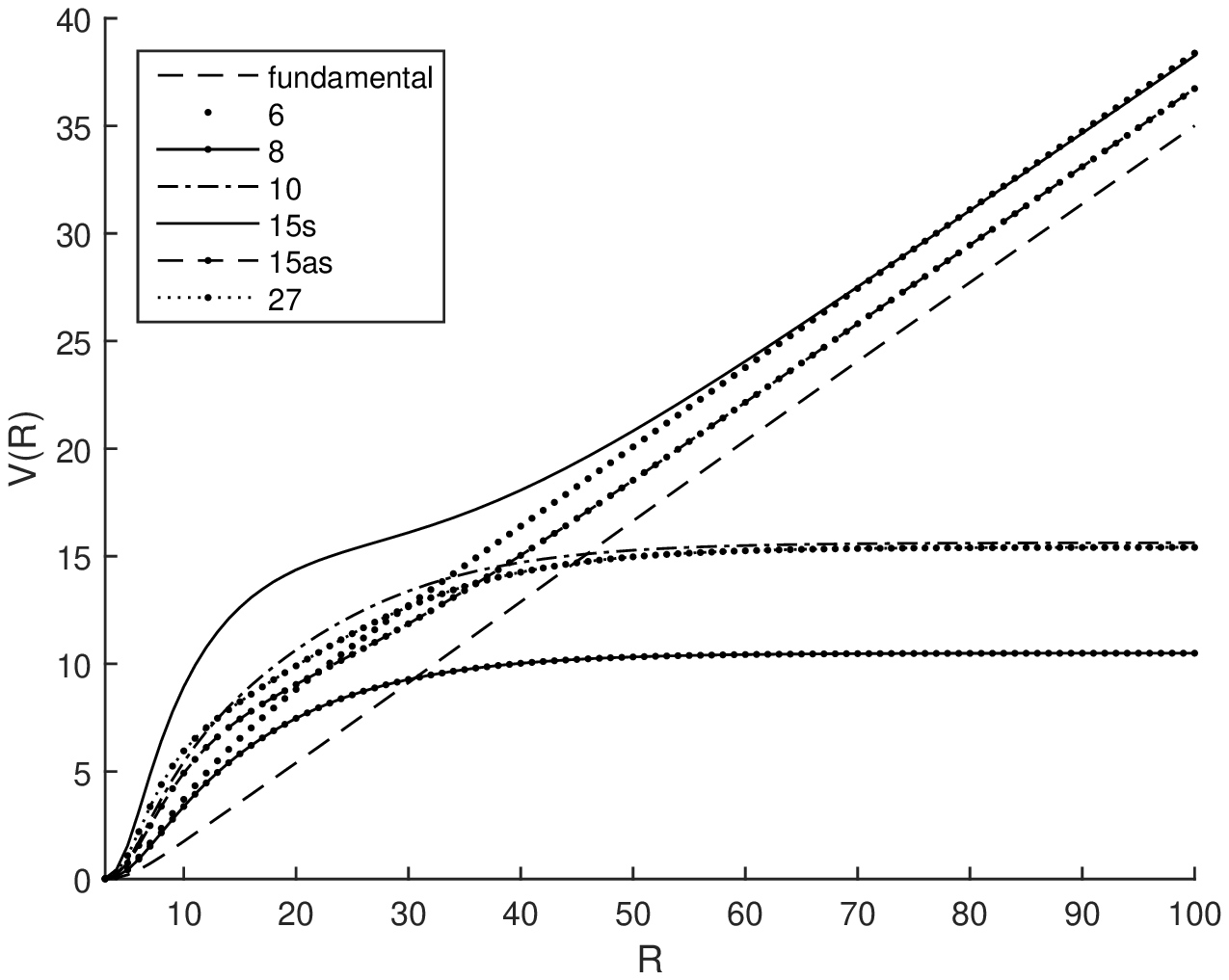}
\caption{ 
Potentials between static sources of $ SU(3)$ gauge group using thick center vortex model. At some intervals, specially for representation $15_s$  at the distances $10 \leqslant R  \leqslant  30$ a non physical concavity is observed.}
\label{su3tak}
\end{center}
\end{figure*}

In the $SU(3)$ gauge group the generators of the fundamental representation are Gell Mann matrices. In $SU(3)$ there are two diagonal matrices and they are used in this model. The normalization conditions are applied for each representation. Also diagonal generators of higher representations are obtained using the tensor method. Using the thick center vortex model the quark potentials behaviour is shown in different representations of the $SU(3)$ group in Figure \ref{su3tak}. Again the true $3$-ality is observed in different representations of the group. For the adjoint  representation with $0$-ality a screening asymptotic behaviour is observed. For the other representations asymptotic linear behaviour is observed. Again an upward concavity is observed in the quark potential using this model in the $SU(3)$ gauge theory at distances $R=10$ to $40$. 
This non-physical behaviour should be removed from the model. Previously, some methods are used to remove this non-physical behaviour of the quark potential of the thick center vortex model \cite{Rafibakhsh2010}. We would like to understand more about the reason of appearance of such behaviour in the quark potentials. Due to this reason, a close look at the concavity criteria is done in the next section.
  
\section{A proof for the concavity of quark potential}
The force between quark anti quark should be attractive and this lead a monotone concave quark anti-quark potential relative to the distance. To understand the reason, there is a simple proof based on the Schwarz type inequality \cite{bachs}. Consider a hyper cubic lattice with sites $p=(p^1,p^2,p^3,p^4) \in Z^4$. A $U(p,p^{\prime})$ gauge field can be introduced from one point $p$ to the next point $p^{\prime}$. For a path like $ \Omega=({p}_0,{p}_1,{p}_2,{p}_3,{p}_4,\ldots,{p}_f)$, $U(\Omega)$ can be obtained as
 
\begin{equation}
U(\Omega)=U({p}_0,{p}_1)U({p}_1,{p}_2)U({p}_2,{p}_3)\ldots U({p}_{f-1},{p}_f)
\end{equation}
Also for an opposite path $-\Omega=p_{f},p_{f-1},\ldots,p_2,p_1, p_0 $ the following condition is satisfied:
\begin{equation}
U(-\Omega)=U^{\dagger}(\Omega)
\end{equation}
  Consider the action as 
\begin{equation}
S=\frac{1}{g^2}\sum_{ plaquettes  p}ReTrU(p)
\end{equation}
Then the static quark antiquark potential can be obtained with a long rectangle Wilson loop in the $R \times T$ plane when $T\rightarrow \infty $    : 
\begin{equation}
V(R)=\lim_{T\rightarrow \infty}\lbrace - \frac{1}{T}ln\langle tr U(W)\rangle + const. \rbrace
\end{equation}
Where
\begin{equation}
\langle tr U(W)\rangle=\frac{\int \prod_P [dU(P)]e^{-S}tr U(W)}{\int \prod_P [dU(P)]e^{-S}}
\end{equation}

\begin{figure*} [!ht]
\begin{center}
\includegraphics [scale=0.6] {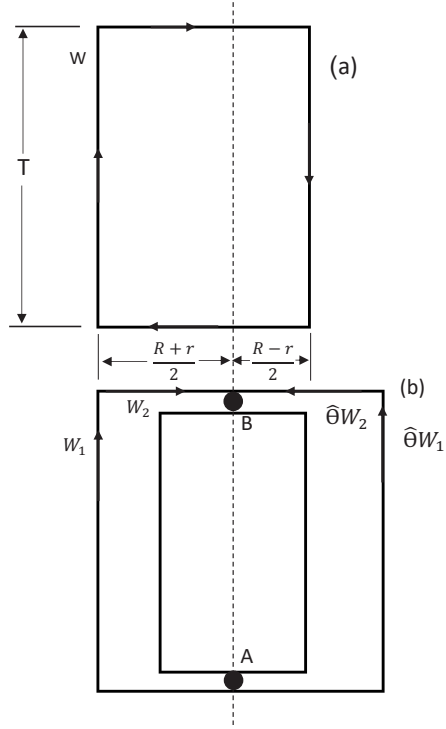}
\caption{(a) The large Wilson loop W, with sides $R\times T$. The dashed line is its intersection with the reflection hyper plane. (b) The paths $ W_1$  and  $W_2$, going from A to B, used in inequality \ref{tr u}, and their reflections. Note that W is the combination of $W_1$  and  $-\hat{\Theta}W_2$. The dashed line can be an arbitrary line crossing the Wilson loop in the plane.} 
\label{proof1}
\end{center}
\end{figure*} 

$dU(P)$ is the group measure. $Q_Z=\int \prod_P [dU(P)]e^{-S}$ is used for the further calculation.The action has reflection positivity which is due to a positive metric Hilbert space. To understand this, it is better to consider a three dimensional hyper plane normal to the primary axis of the lattice. For example consider $p^1=0$ hyper plane and denote the sites of links and plaquettes that lie above, on, below with $L_{+}, L_{0}, L_{-}$   respectively. Consider the border place is arbitrary and due to the $p^1$, consider a $\hat{\Theta}$ reflection operator such that 

\begin{eqnarray}
\hat{\Theta} f(U(P))=f^*(\hat{\Theta} U(P))=f^*(U(\hat{\Theta}P))\\
\hat{\Theta} p =\hat{\Theta} (p^1,p^2,p^3,p^4)=(-p^1,p^2,p^3,p^4),\\
\hat{\Theta} P=\hat{\Theta} (p ,{p^{\prime}})=(\hat{\Theta} p,\hat{\Theta} p^{\prime})
\end{eqnarray}

$\hat{\Theta}$ is called Osterwalder-Schrader positivity or reflective positivity \cite{Osterwalder} which is the main ingredient for establishing the existence of a positive semidefinite self-adjoint Hamiltonian.
 Using this operator it is possible to write all functions $f$ on $L_0 \cup L_+ $  as 
 
\begin{eqnarray} \label{schwarzinequality}
\langle f \hat{\Theta} f\rangle =Q_Z^{-1}\int \prod _{P \in L_0} [dU(P)]exp(-\frac{1}{g^2}\sum_{p\in L_0}tr U(p)) \nonumber \\
 \int \prod _{P \in L_+}[dU(P)] f(U(P))exp(-\frac{1}{g^2}\sum_{p\in L_+}tr U(p))\nonumber \\
  \int \prod _{P \in L_-}[dU(P)] f^*(U(\hat{\Theta}P))exp(-\frac{1}{g^2}\sum_{p\in L_-}tr U(p))\nonumber \\ 
 =Q_Z^{-1}\int \prod _{P \in L_0} [dU(P)]exp(-\frac{1}{g^2}\sum_{p\in L_0}tr U(p))\nonumber \\
  \vert \int \prod _{P \in L_+}[dU(P)]f(U(P))
 exp(-\frac{1}{g^2} \sum_{p\in L_+} tr U(p)) \vert ^2 \geq 0 
\end{eqnarray}

In which Schwarz-type inequality $\langle f_1 \hat{\Theta} f_2 \rangle^2 \leq \langle f_1 \hat{\Theta} f_1 \rangle$ $\langle f_2 \hat{\Theta} f_2 \rangle$ is used. Consider  this inequality for the hyper plane parallel to the time axis and normal to the  Wilson loop in figure \ref{proof1}. Then using the reflection properties it leads to:
 
\begin{eqnarray}\label{tr u}
\langle tr U(W) \rangle=\sum_{i,j} \langle U(W_1)_{ij} \hat{\Theta}U(W_2)_{ij} \rangle\\ 
\langle tr U(W) \rangle\leqslant \sum_{i,j} \langle U(W_1)_{ij} \hat{\Theta}U(W_1)_{ij} \rangle^{\frac{1}{2}}  \langle U(W_2)_{ij} \hat{\Theta}U(W_2)_{ij} \rangle ^{\frac{1}{2}} \nonumber \\ 
\langle tr U(W) \rangle\leqslant \langle tr [U(W_1) U(-\hat{\Theta}W_1)] \rangle^{\frac{1}{2}}  \langle tr [U(W_2) U(-\hat{\Theta}W_2)] \rangle^{\frac{1}{2}}\nonumber
\end{eqnarray} 
Using the definition for the quark potential this inequality means
 \begin{equation}\label{V(R)}
V(R)\geqslant \frac{1}{2} V(R-r) + \frac{1}{2} V(R+r) 
\end{equation} 
This means a concave quark potential. So it seems the reflection symmetry for the functions and variable is the building block of concavity criteria and if in a model such reflection properties is not considered it leads to the violation of the concavity condition. In the thick center vortex idea a vortex flux is considered which is not reflective relative to the an arbitrary line. So, to obey the concavity in the model it seems such arbitrary symmetry is essential. In the next section it is shown that such symmetry is not considered for the thick center vortex flux in plane and try to consider reflective symmetric vortex fluxes relative to an arbitrary line and study its effects on the quark potential.
 
\section{Elimination of the concavity using  an  arbitrary symmetric vortex flux in the space time hyper plane }
\begin{figure*} [!ht]
\includegraphics [scale=0.6] {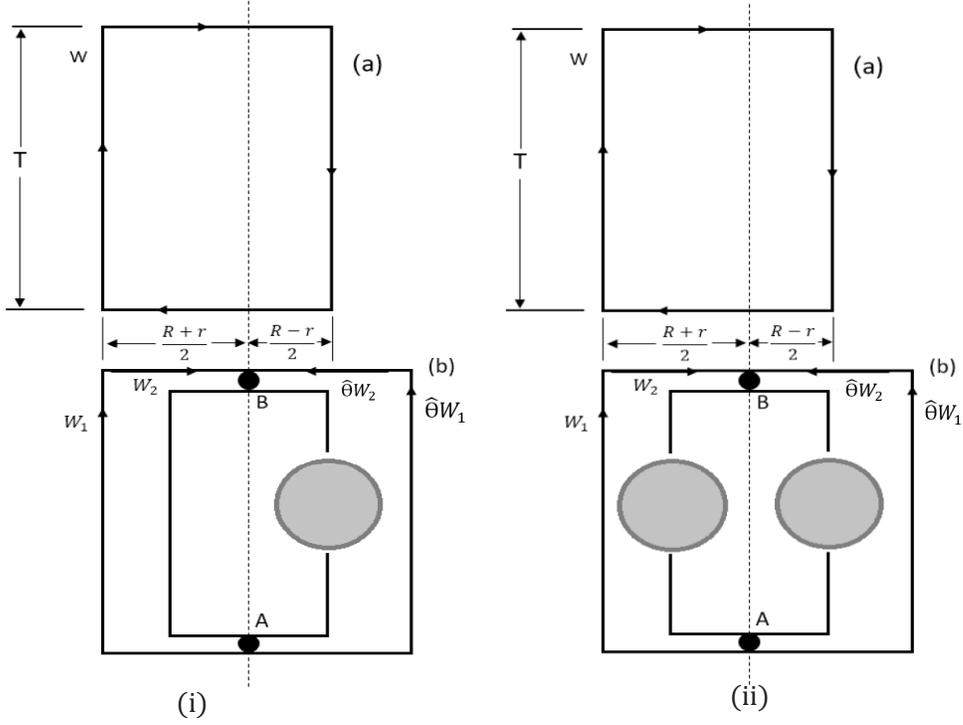}
\caption{(i) The effects of overlapping a vortex flux with the Wilson loop. As it can be seen, the arbitrary symmetry in the space time hyperplane is broken by the vortex flux. (ii) A symmetric vortex flux is needed to establish the symmetry in the space time hyper plane. }
\label{hyperplane}
\end{figure*}

In the previous section a proof for the concavity is explained. A symmetry in the space time hyper plane (figure 3) for the Wilson loop is essential to obtain this criterion. However, considering the presence of the thick center vortices in the plane breaks this symmetry. Then for such asymmetry the formulas (\ref{schwarzinequality},\ref{tr u},\ref{V(R)}) cannot lead to the concavity criterion. The question here is how we can use the concavity proof again in the presence of the vortices in the plane? Figure \ref{hyperplane}.i shows the situation of the presence of the vortices in the plane. As it is clear the vortex flux in the plane breaks the arbitrary symmetry of the plane. To investigate this, we take a look at the center vortex mechanism and try to apply the changes to the thick center vortex model related to the concavity proof. 

In the center vortex mechanism, the vortices effects on a Wilson loop is considered through the center elements of the group $Z$. No thickness is considered for the vortices in this model. This is clear in the equation \ref{W}. However, to consider the concavity proof for the center vortex model the symmetry of piercing of the vortices should be accounted because of the presence of the $Z$ in the upper or lower plane. This can be done through considering two vortex piercing, one in the lower plane and one in the upper plane. The additional vortex leads to a symmetry for the Wilson loop in the upper and lower plane. A $Z^*$ is considered for the presence of a symmetric vortex relative to the $Z$ relative to the reflection operator $\hat{\Theta}$. This is the simplest ansatz one can consider. Due to this:
 
 \begin{eqnarray} \label{schwarzinequality1}
\langle f \hat{\Theta} f\rangle =Q_Z^{-1}\int \prod _{P \in L_0} [dU(P)]exp(-\frac{1}{g^2}\sum_{p\in L_0}tr U(p))\nonumber \\
\int \prod _{P \in L_+}[dU(P)] U_1U_2Z\ldots U_l\exp(-\frac{1}{g^2}\sum_{p\in L_+}tr U(p))\nonumber \\
\times  \int \prod _{P \in L_-}[dU(P)] U^\dagger_l\ldots Z^*U^\dagger_2U^\dagger_1 exp(-\frac{1}{g^2}\sum_{p\in L_-}tr U(p)) \nonumber\\
 =Q_Z^{-1}\int \prod _{P \in L_0} [dU(P)]exp(-\frac{1}{g^2}\sum_{p\in L_0}tr U(p))\\
  \vert \int \prod _{P \in L_+}[dU(P)]U_1U_2Z\ldots U_l exp(-\frac{1}{g^2}\sum_{p\in L_+}tr U(p))\vert ^2 \geq 0 \nonumber
\end{eqnarray} 
 
again the Schwarz inequality is applied. $ZZ^*=|Z|^2$ can be omitted:
\begin{equation}
|Z|^4\langle f_1 \hat{\Theta} f_2 \rangle^2 \leq |Z|^2\langle f_1 \hat{\Theta} f_1 \rangle |Z|^2\langle f_2 \hat{\Theta} f_2 \rangle
\end{equation}
 Again, all of the previous proof of the concavity becomes valid. So, in the center vortex model considering a symmetric ansatz for the piercing of the vortex in the Wilson loop can explain the situation. 
 
For the thick center vortex $W(C)=Tr[UUU...U ]\longrightarrow Tr[UU....(G(x,s))U]$ is used for the piercing of the vortex. $G(x,s)$ is introduced in equation \ref{G}. The validity of the concavity should be examined. Here a vortex ansatz with a symmetric $G$ ansatz is considered. Due to the vortex thickness a $G_1$ portion is considered for the presence of the vortex in the lower plane and a $G_1^*$ portion is considered for the piercing of the vortex in the upper plane. Also consider $G_1(x,s) G_1^*(\hat{\Theta} x,s)=G$. So it seems that relative to the border, the ansatz divide the vortex flux in to the two symmetric parts one in the upper  plane and one in the lower plane:

\begin{eqnarray} \label{schwarzinequality2}
\langle f \hat{\Theta} f\rangle =Q_Z^{-1}\int \prod _{P \in L_0} [dU(P)]exp(-\frac{1}{g^2}\sum_{p\in L_0}tr U(p))\nonumber\\
\int \prod _{P \in L_+}[dU(P)] U_1U_2G_1\ldots U_l\exp(-\frac{1}{g^2}\sum_{p\in L_+}tr U(p)) \nonumber \\
\times  \int \prod _{P \in L_-}[dU(P)]U^\dagger_l\ldots G_1^*U^\dagger_2U^\dagger_1 exp(-\frac{1}{g^2}\sum_{p\in L_-}tr U(p)) \nonumber\\
 =Q_Z^{-1}\int \prod _{P \in L_0} [dU(P)]exp(-\frac{1}{g^2}\sum_{p\in L_0}tr U(p))\\
  \vert \int \prod _{P \in L_+}[dU(P)]UUG^{\frac{1}{2}}UU\ldots exp(-\frac{1}{g^2}\sum_{p\in L_+}tr U(p))\vert ^2 \geq 0 \nonumber
\end{eqnarray}   
Again, the previous proof can be true by considering an ansatz with symmetric $G$ relative to the border of the upper and lower plane. So, if the symmetry restored, the concavity proof would be applicable again. A symmetric ansatz with the equal portion relative to the border of the upper and lower plane leads again to the validation of concavity formulas. Figure \ref{hyperplane}.ii shows this situation in which a symmetric vortex profile with equal distances to the intersection is considered. To consider such situation, two vortex fluxes are introduced as 

 \begin{eqnarray}
\vec{\alpha}_{1}(x)=\vec{N}^n[1-tanh(ay(x+R)+\frac{b}{R})]\nonumber\\
\vec{\alpha}_{2}(x)=\vec{N}^n[1-tanh(ay(x-R)+\frac{b}{R})]\nonumber\\
\vec{\alpha}(x)=\vec{\alpha}_{1}(x)+\vec{\alpha}_{2}(x)
\end{eqnarray}

\begin{figure*} [!ht]
\begin{center}
\includegraphics [scale=0.8] {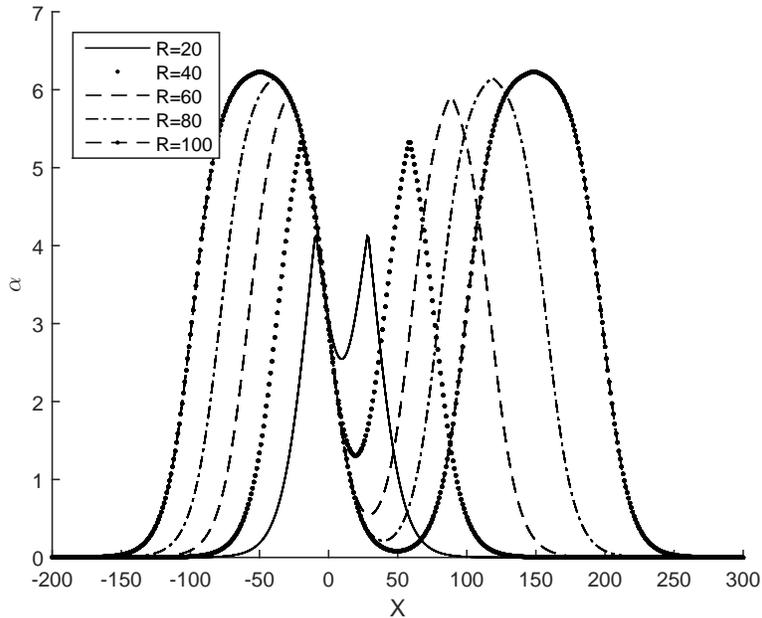}
\caption{Two symmetric vortex fluxes are considered to restore the arbitrary symmetry of the space time hyper plane for the intervals of x+R and x-R in the $SU(2)$ gauge group.}
\label{sharttwo}
\end{center}
\end{figure*}

\begin{figure*} [!ht]
\begin{center}
\includegraphics [scale=0.8] {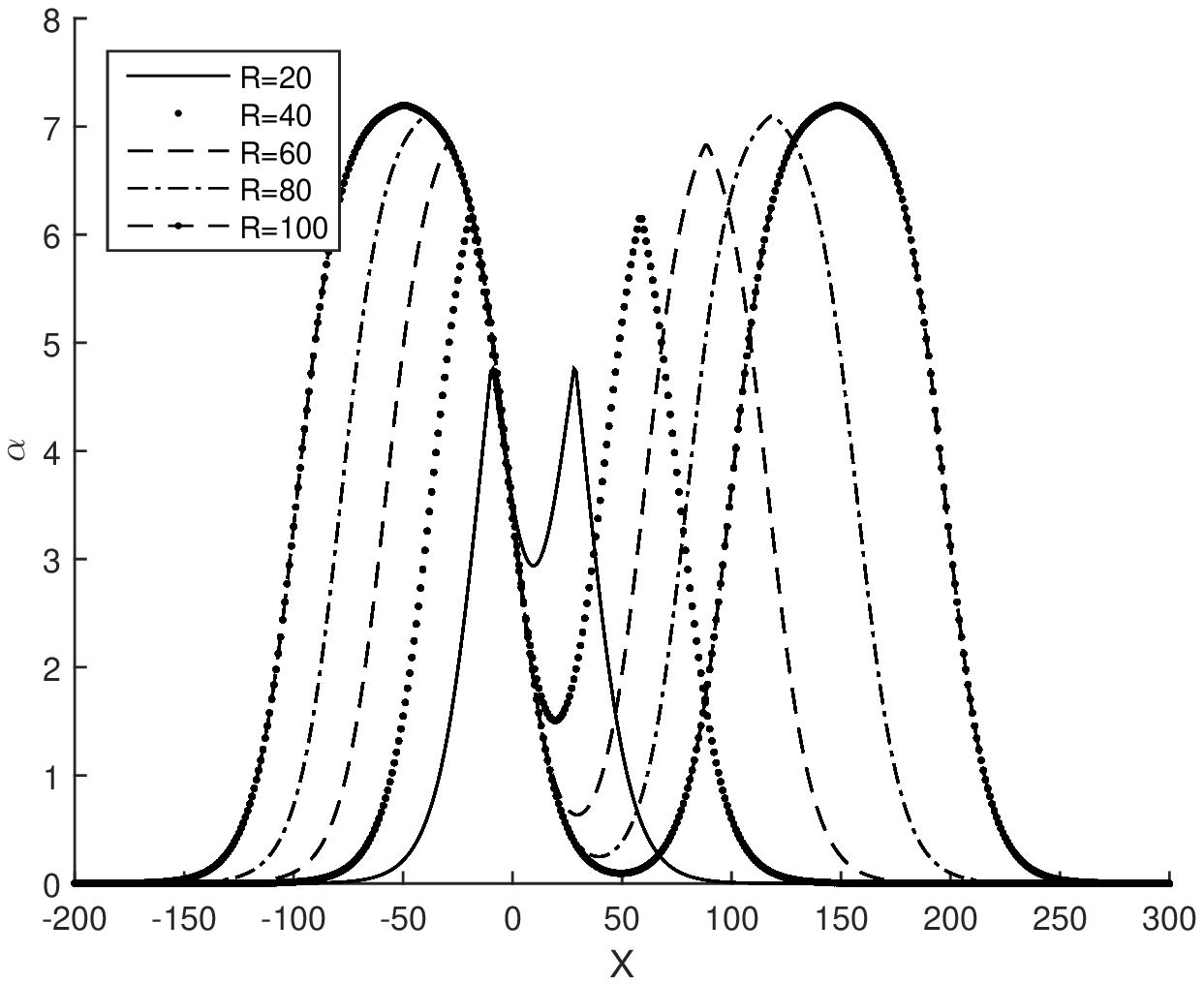}
\caption{Two symmetric vortex fluxes are considered to restore the arbitrary symmetry of the space time hyper plane for the intervals of x+R and x-R in the $SU(3)$ gauge group.}
\label{shartwosu3}
\end{center}
\end{figure*}

\begin{figure*} [!ht]
\begin{center}
\includegraphics [scale=0.8] {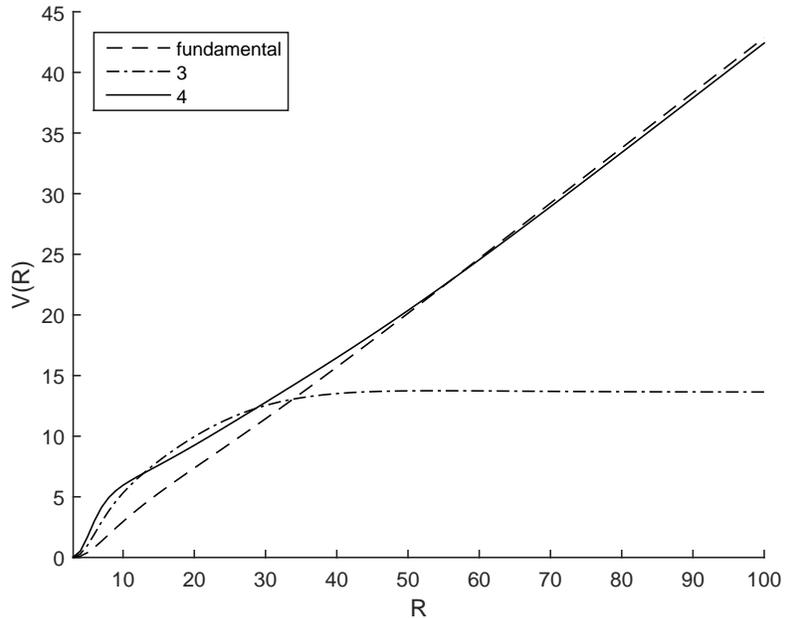}
\caption{Quark potentials behaviour for the $SU(2)$ gauge group and its various representations are plotted with the introduced arbitrary symmetric flux. By comparing this figure and Figure \ref{su2tak}, it is well seen that the non-physical concavity especially for representation $4$ is reduced.}
\label{su2two}
\end{center}
\end{figure*}

\begin{figure*} [!ht]
\begin{center}
\includegraphics [scale=0.8] {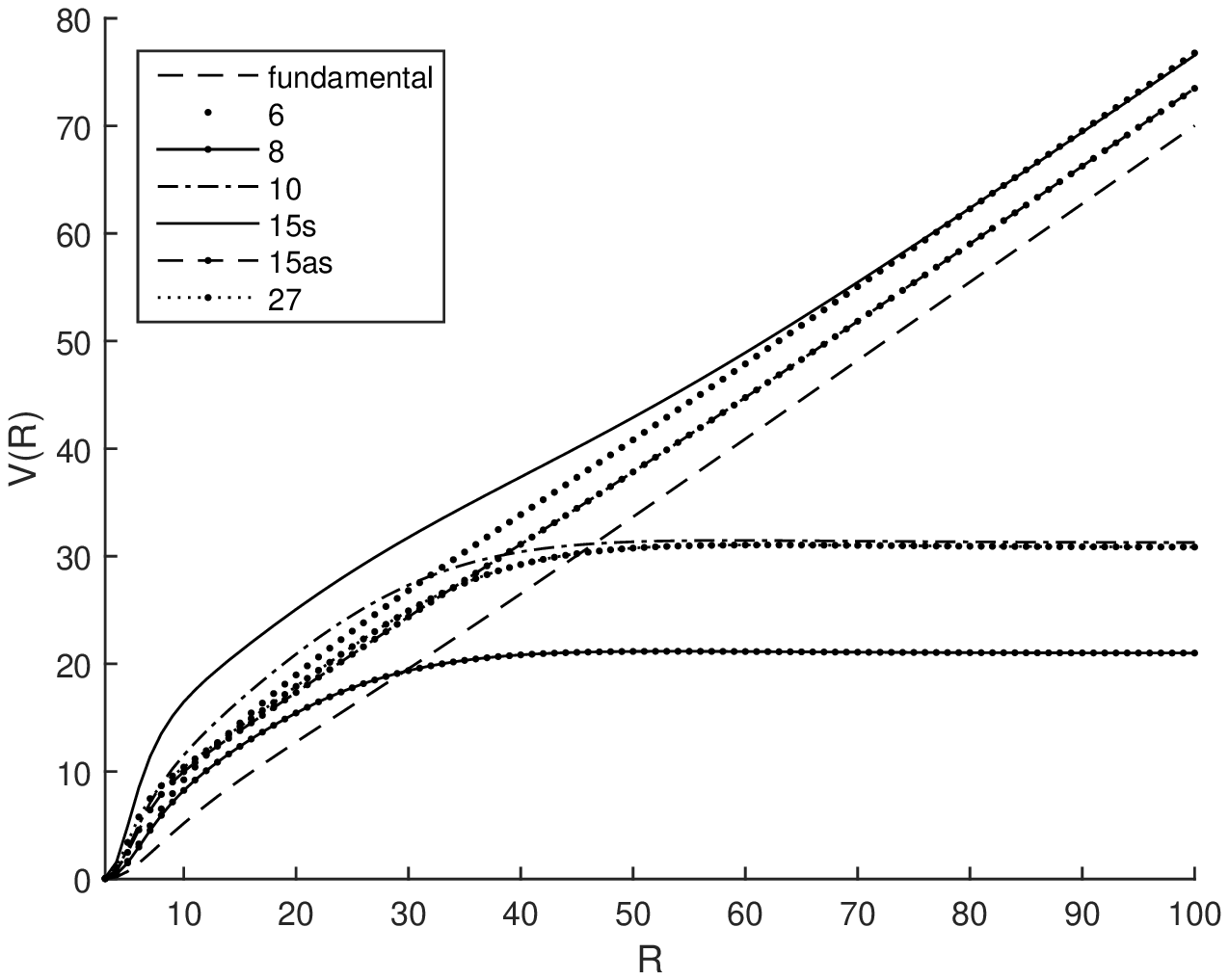}
\caption{Quark potentials behaviour for the $SU(3)$ gauge group and its various representations are plotted with the introduced arbitrary symmetric flux. By comparing this figure and Figure \ref{su3tak}, it is well seen that the non-physical concavity especially for representation $15_s$ is completely eliminated in the intermediate region and also an overall concavity reduction is observed.}
\label{su3two}
\end{center}
\end{figure*}

$R$ is the space side of the Wilson loop. Using such ansatz the symmetry is restored and the concavity proof becomes valid.  For the symmetric flux in the lower plane a vortex with overlapping with the dual Wilson loop ($W_2 \cup \hat{\Theta}W_1$) is considered. Figures \ref{sharttwo} and \ref{shartwosu3} shows the behaviour of the vortex flux relative to the center of the vortices. Figure \ref{su2two} shows the quark potentials for the different representation of the $SU(2)$ group using the symmetric vortex flux. Figure \ref{su3two} shows the quark potentials for the representation of the $SU(3)$ group. As it is seen the non-physical concavity present in figures \ref{su2tak} and \ref{su3tak} are removed in figures \ref{su2two} and \ref{su3two} in the Casimir region due to the restoration of the an arbitrary symmetry in the plane. So to obtain the true quark potentials using the thick center vortex model, a symmetric vortex flux can be introduced in the plane. The Casimir region of the potential is the part of the potential which is due to the vortex thickness. Considering symmetric vortex thickness has been removed the concavity from this region and an overall reduction in the concavity is seen.

Note that the line for considering such symmetry is arbitrary line which can be considered in the vacuum. There is no forced place for presence of such border in the vacuum. Especially the line is considered somewhere with no symmetry relative to the Wilson loop to show such arbitrariness of the position of the line or border between upper and lower hyperplane. The line can be considered randomly or for every vortex in the QCD vacuum we can find at least one vortex with similar flux present in the QCD Vacuum. The line in two dimensions or hyperplane in 4 dimensions can be considered with equal distance between these two vortices. For every Wilson loop with the splitting line the proof is correct. For example, if we consider another Wilson loop in the vacuum space again, we can consider such arbitrary line which divide the space of the vacuum in to the upper and lower space. This arbitrariness of the position of the line exclude any new symmetry for the QCD vacuum. But relative to any splitting line we consider a symmetric vortex flux. Due to the fluctuation of the vortex fluxes this consideration is not very far reaching. Instead of considering any fluctuation of the vortex fluxes, a symmetric vortex fluxes are considered for the vortices in the vacuum relative to this arbitrary line. 
Also, we do not consider vortices position fix in the space-time and considering a movement of the vortices by enlarging the Wilson loop. This is done by considering the center of vortex fluxes at $x-R$ and $x+R$ in which $R$ is the space side of the Wilson loop. For example, by expanding the Wilson loop from $R=10$ to $R=100$ the position of these vortex fluxes moves. The movement of the vortex fluxes with such function can be a simple model for the fluctuation of the vortices within the QCD vacuum for this model. So, the symmetry applied in the article is not a real QCD vacuum symmetry but an accidental symmetry in the QCD vacuum due to the fluctuation and movement of vortices in the QCD vacuum. 
 
 Also the domain structure idea for the QCD vacuum can be examined \cite{Faber1998,Lookzadeh2012}. In the domain structure picture of the QCD vacuum, trivial center element of the group also can affect the behaviour of the linear potential. So in this model which is a modification of the thick center vortex model, the trivial center element portion is considered in the model through the quark potential equation as: 
\begin{eqnarray}
V(R)=\sum_{x}Ln\lbrace 1-\sum_{n=0}^{N-1}f_{n}(1-Reg_{r}[\vec{\alpha}_{C}^{n}(x)]\rbrace.
\end{eqnarray}  
  
\begin{figure*} [!ht]
\begin{center}
\includegraphics [scale=0.8] {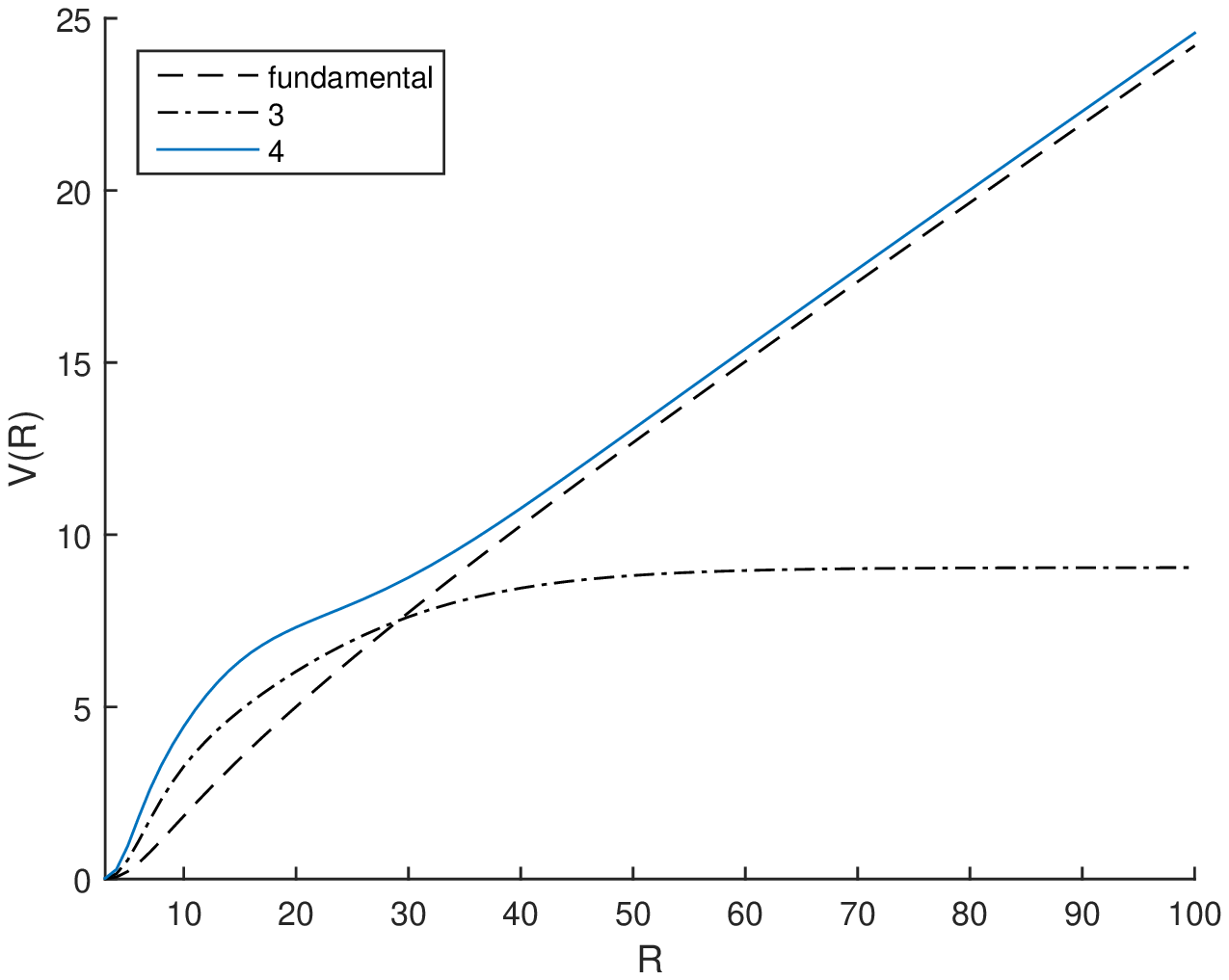}
\caption{Quark potentials behaviour for the $SU(2)$ gauge group and its various representations are plotted using the domain structure model.}
\label{su2onedomain}
\end{center}
\end{figure*}

\begin{figure*} [!ht]
\begin{center}
\includegraphics [scale=0.8] {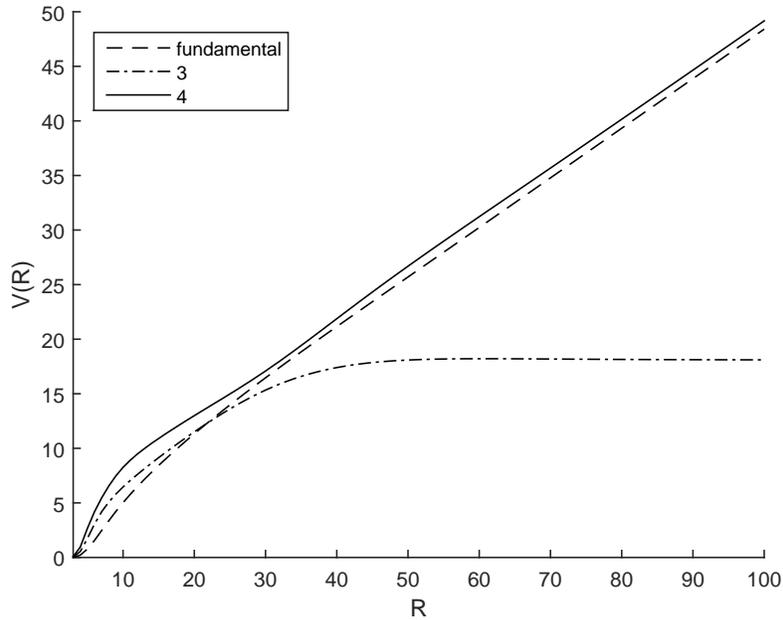}
\caption{Quark potentials behaviour for the $SU(2)$ gauge group and its various representations are plotted using the domain structure model with a symmetric vortex flux. A better reduction of the concavity is observed using the symmetric vortex profile using the domain structure.}
\label{su2twodomain}
\end{center}
\end{figure*}

\begin{figure*} [!ht]
\begin{center}
\includegraphics [scale=0.8] {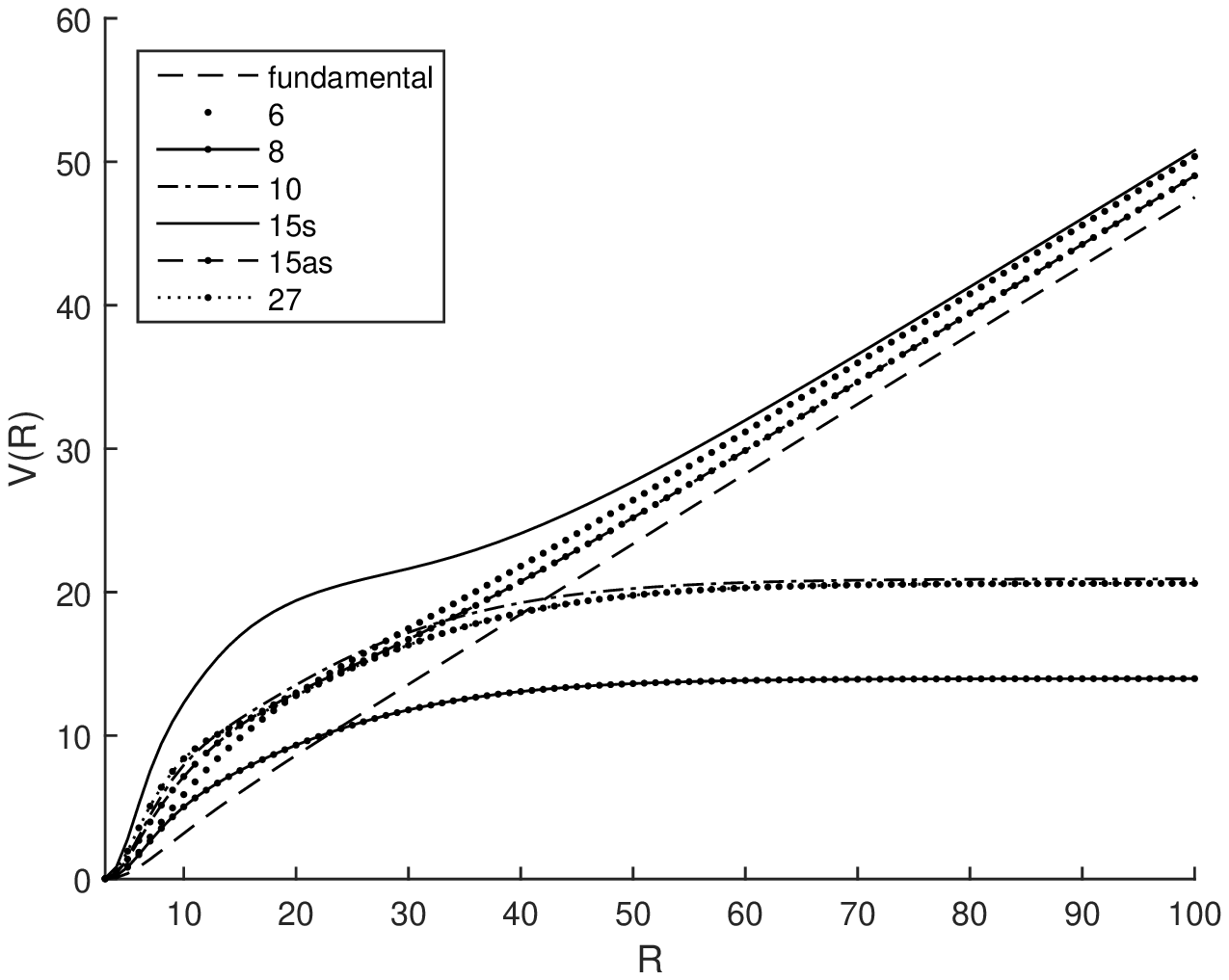}
\caption{Quark potentials behaviour for the $SU(3)$ gauge group and its various representations are plotted using the domain structure model.}
\label{su3onedomain}
\end{center}
\end{figure*}

\begin{figure*} [!ht]
\begin{center}
\includegraphics [scale=0.8] {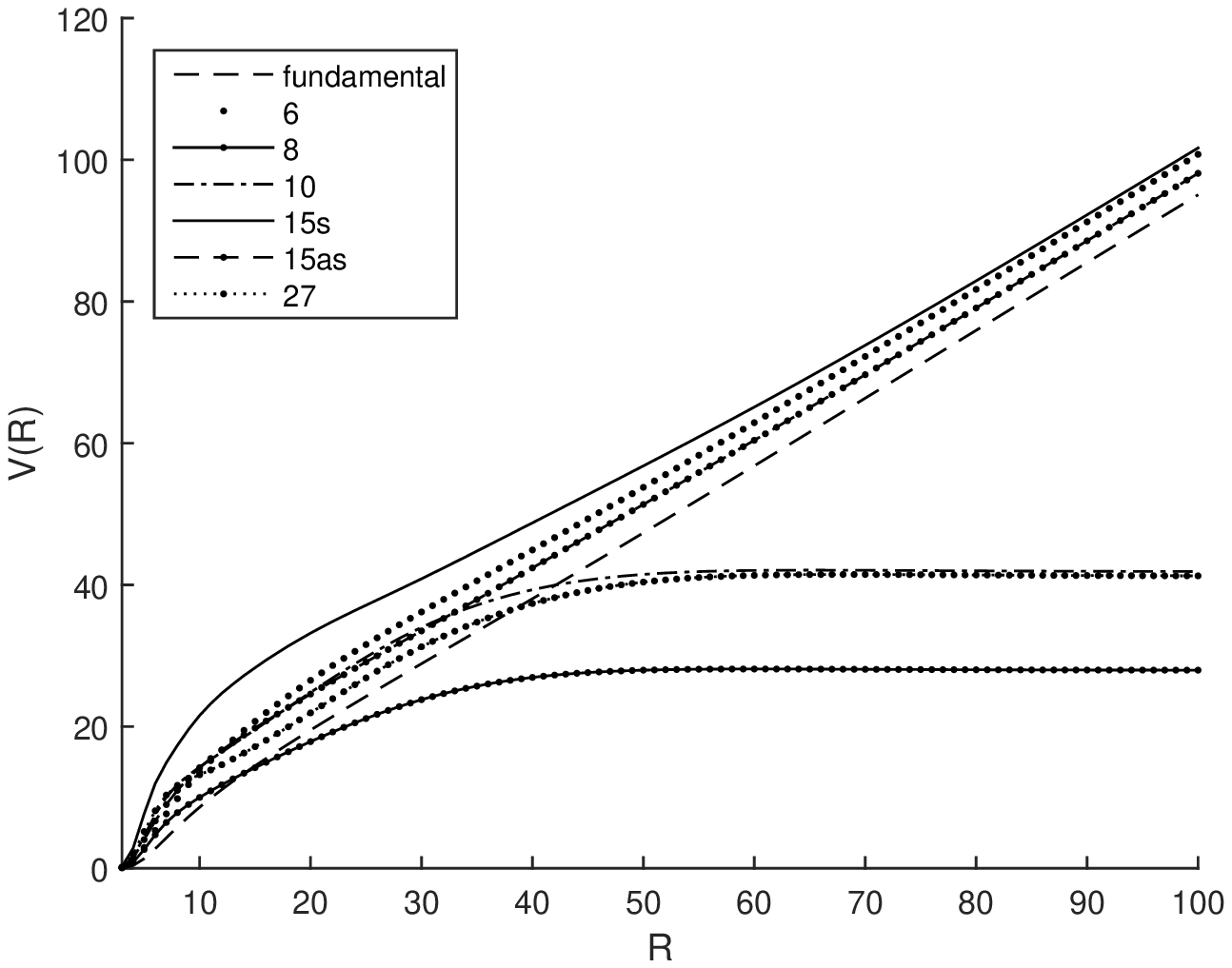}
\caption{Quark potentials behaviour for the $SU(3)$ gauge group and its various representations are plotted with the introduced arbitrary symmetric flux using the domain structure model. By comparing this figure and Figure \ref{su3onedomain}, it is well seen that the non-physical concavity especially for representation $15_s$ is completely eliminated in the Casimir region. A better reduction of an overall concavity is observed relative to the thick center vortex model.}
\label{su3twodomain}
\end{center}
\end{figure*}

$\vec{\alpha}_{C}^{0}(x)$ is the center element portion and its flux is normalized to the trivial center element $I$. $f_0$ is the probability of trivial domain piercing the Wilson loop minimal area. A $f_0=0.025$ is considered for the $SU(2)$ and $f_0=0.05$ is considered for the $SU(3)$ group to obtain the quark potential. Fig. \ref{su2onedomain} and \ref{su3onedomain} shows the quark potential for the $SU(2)$ and $SU(3)$ before considering the symmetric profile ansatz. Again the concavity is present at distances $R=10 to 40$.  Fig \ref{su2twodomain} and \ref{su3twodomain} shows the quark potential for the $SU(2)$ and $SU(3)$ considering the symmetric ansatz using the domain structure model. As it can be seen again the concavity is removed from the Casimir region and even an overall better reduction of the concavity is observed relative to the thick center vortex model. 

\section{Casimir scaling considering the symmetric vortex fluxes}  
\begin{figure*} [!ht]
\begin{center}
\includegraphics [scale=0.8] {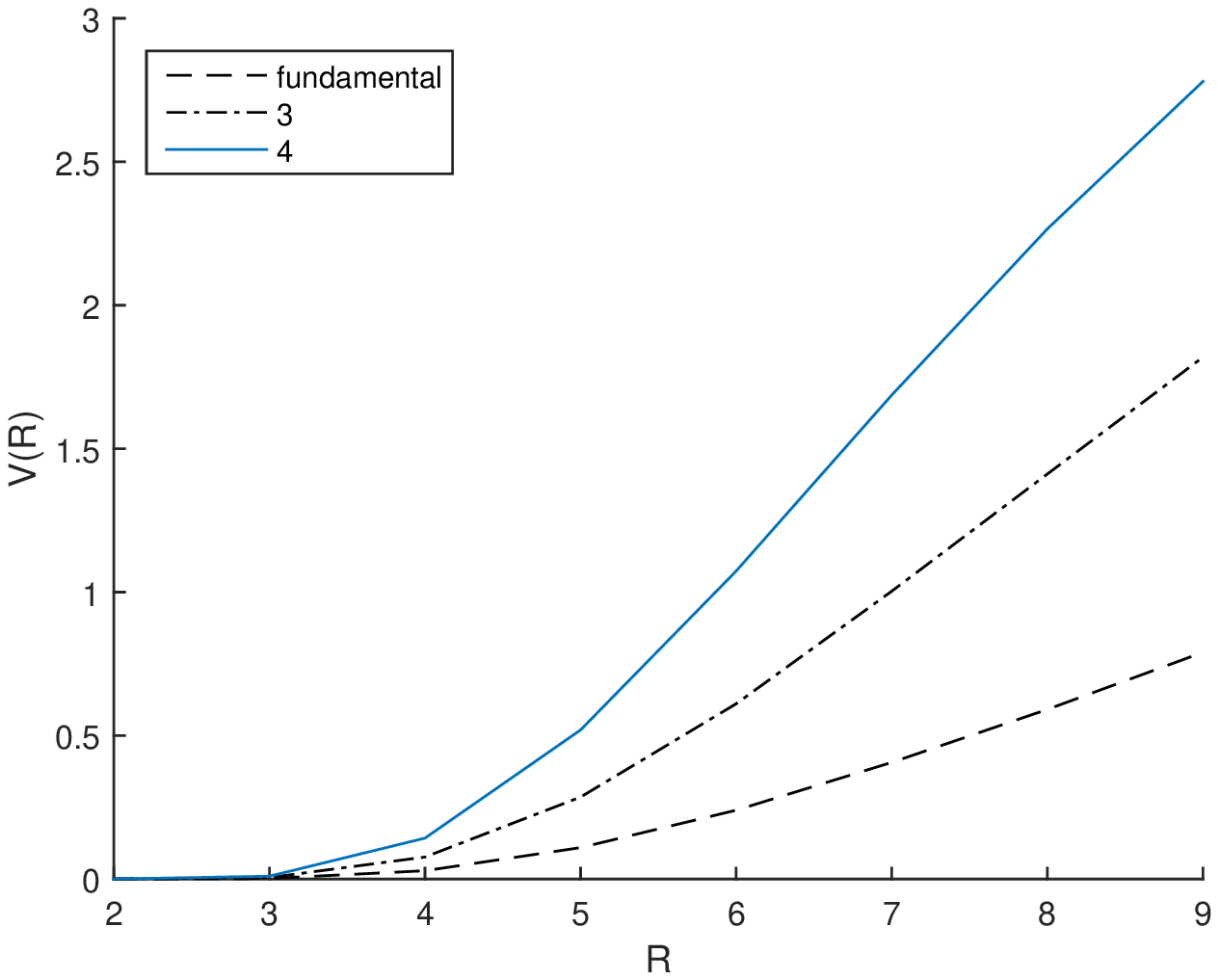}
\caption{Quark potentials behaviour for the $SU(2)$ representations at the intermediate region with non-symmetric vortex fluxes.}
\label{taksu21}
\end{center}
\end{figure*}

\begin{figure} [!ht]
\begin{center}
\includegraphics [scale=0.8] {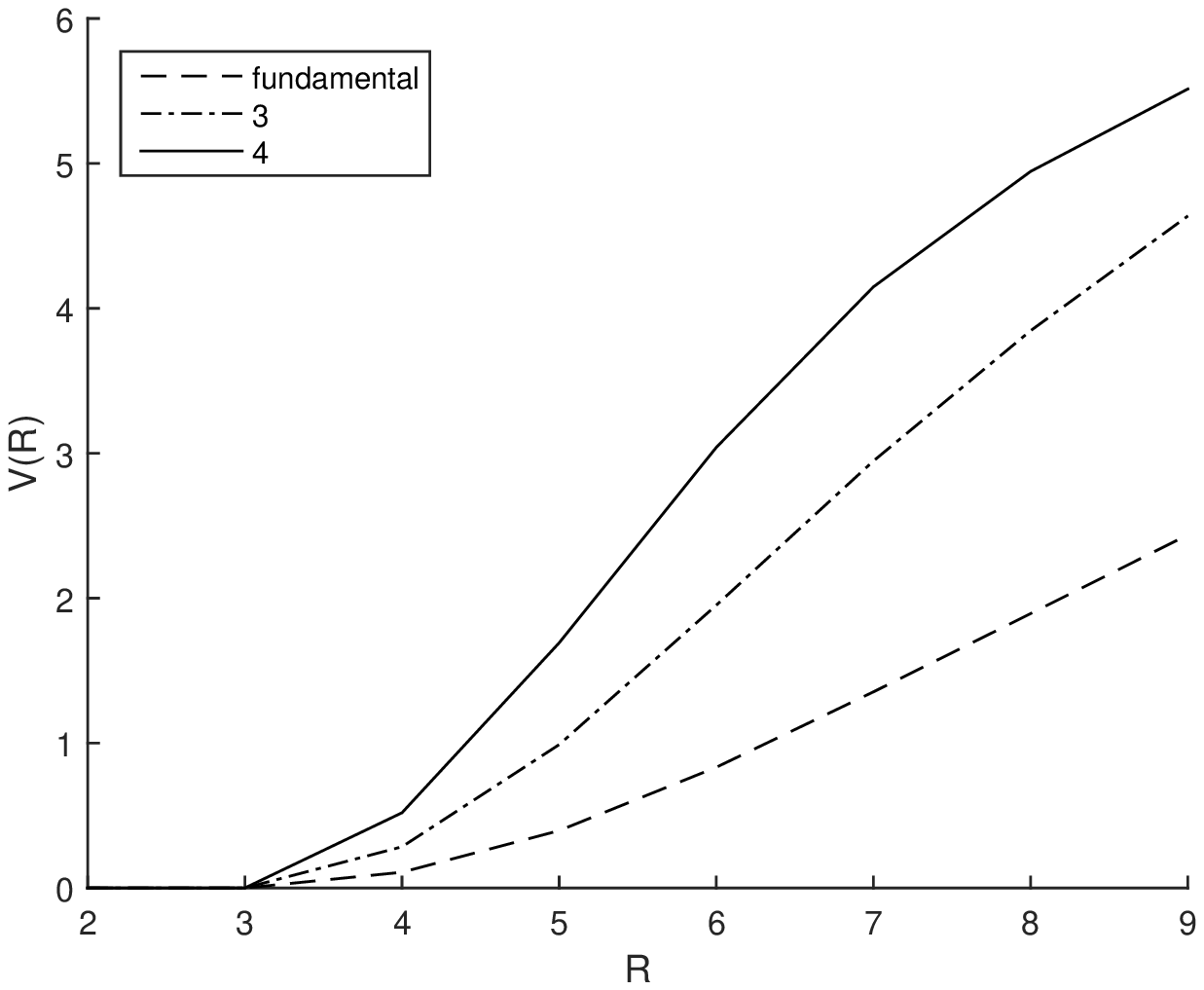}
\caption{Quark potentials behaviour for the $SU(2)$ representations at the intermediate region with arbitrary symmetric vortex fluxes.}
\label{twosu21}
\end{center}
\end{figure}

\begin{figure*} [!ht]
\begin{center}
\includegraphics [scale=0.8] {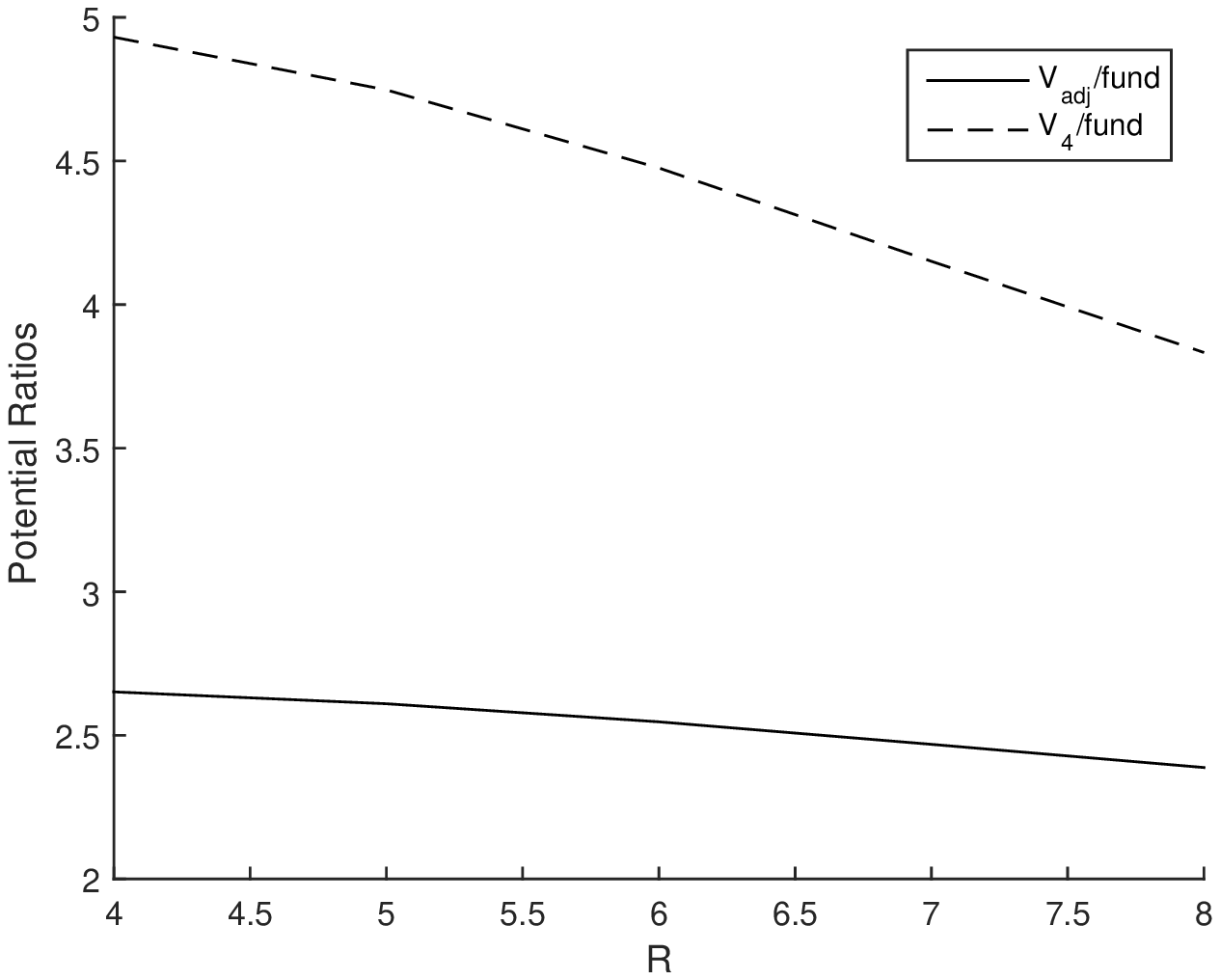}
\caption{Casimir ratios behaviour for the $SU(2)$ representations with non-symmetric vortex fluxes.}
\label{casimirtaksu2}
\end{center}
\end{figure*}

\begin{figure} [!ht]
\begin{center}
\includegraphics [scale=0.8] {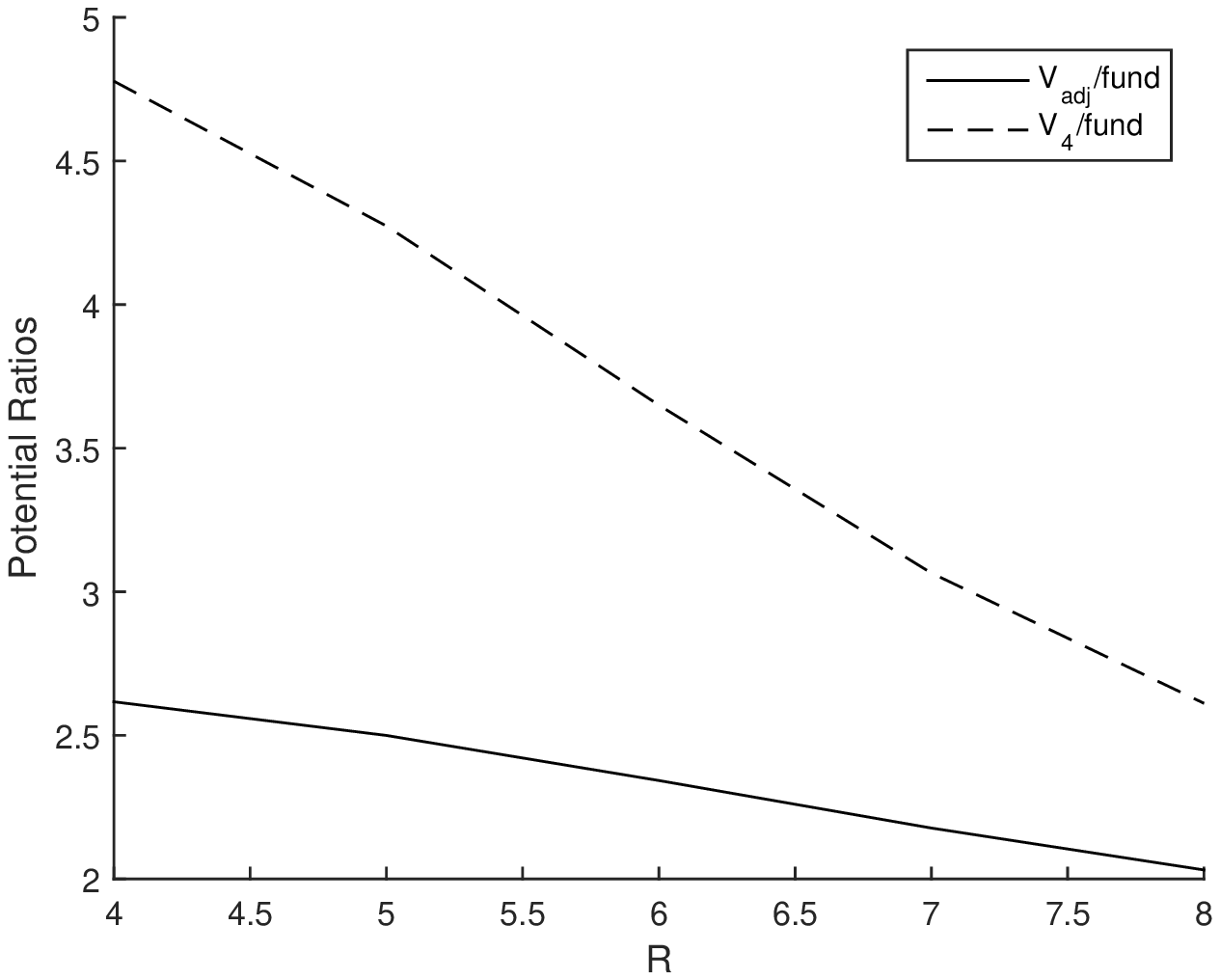}
\caption{Casimir ratios behaviour for the $SU(2)$ representations with arbitrary symmetric vortex fluxes.}
\label{casimirtwosu2}
\end{center}
\end{figure}

\begin{figure*} [!ht]
\begin{center}
\includegraphics [scale=0.8] {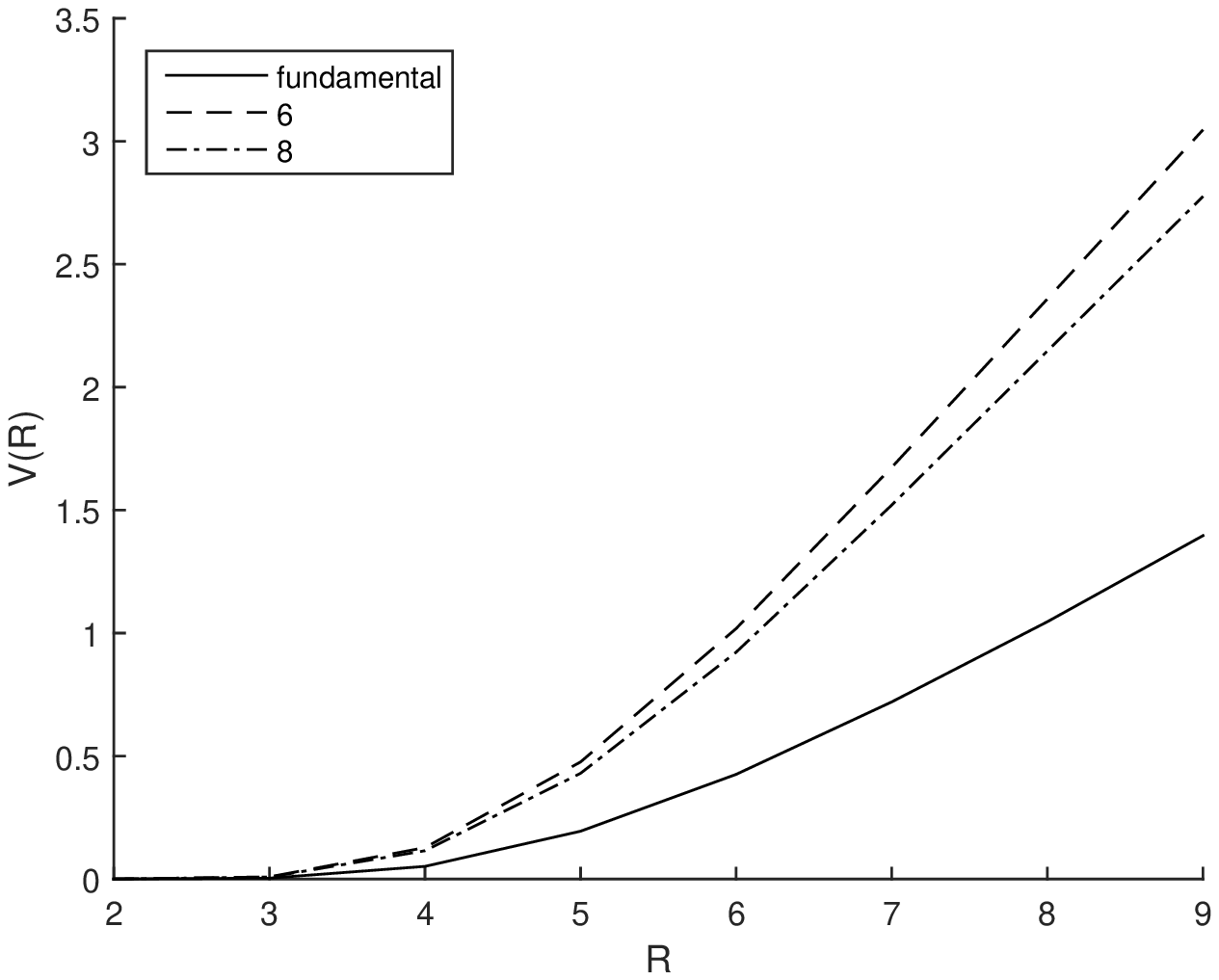}
\caption{Quark potentials behaviour for the $SU(3)$ representations at the intermediate region with non-symmetric vortex fluxes.}
\label{taksu31}
\end{center}
\end{figure*}

\begin{figure} [!ht]
\begin{center}
\includegraphics [scale=0.8] {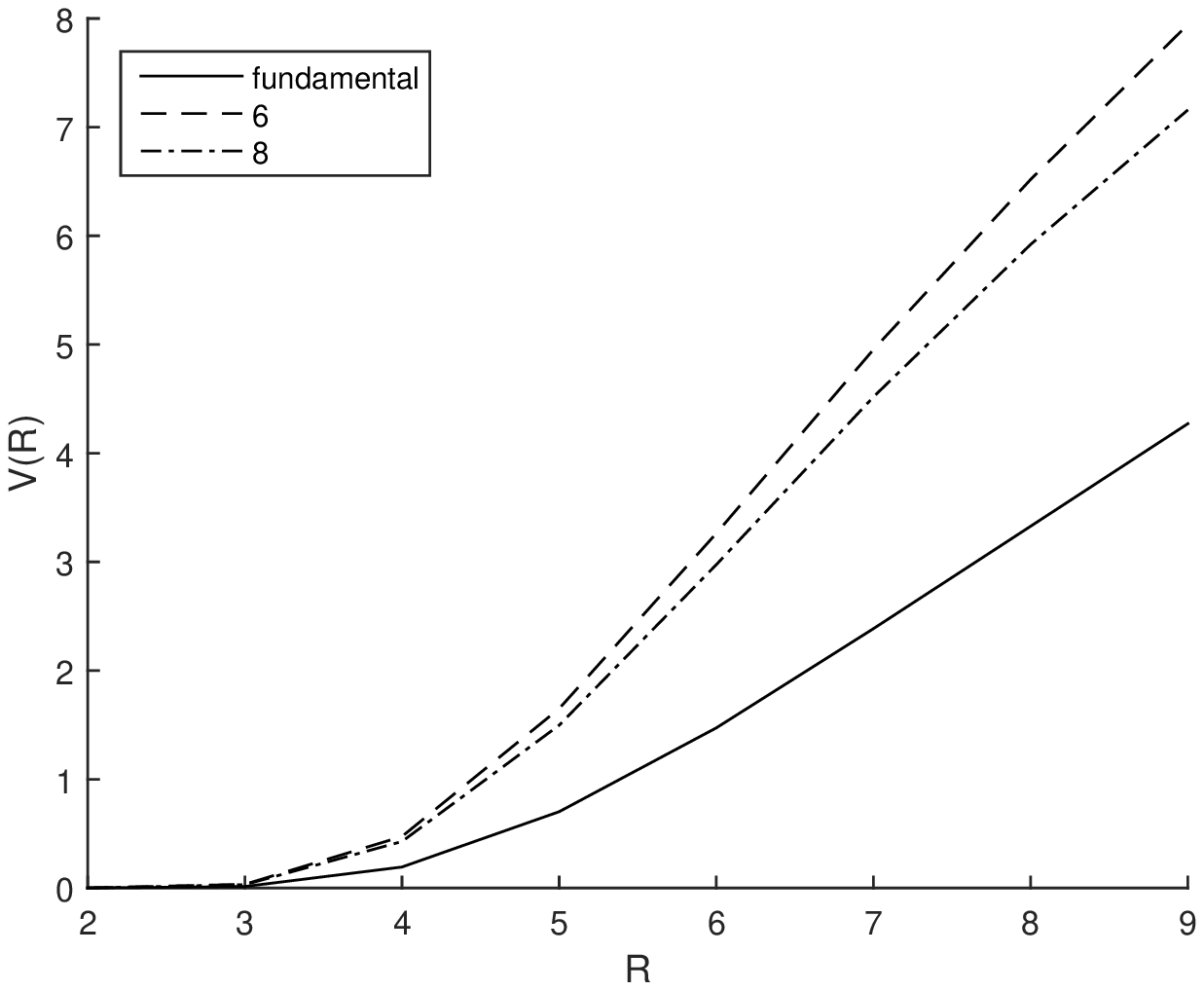}
\caption{Quark potentials behaviour for the $SU(3)$ representations at the intermediate region with arbitrary symmetric vortex fluxes.}
\label{twosu31}
\end{center}
\end{figure}

\begin{figure*} [!ht]
\begin{center}
\includegraphics [scale=0.8] {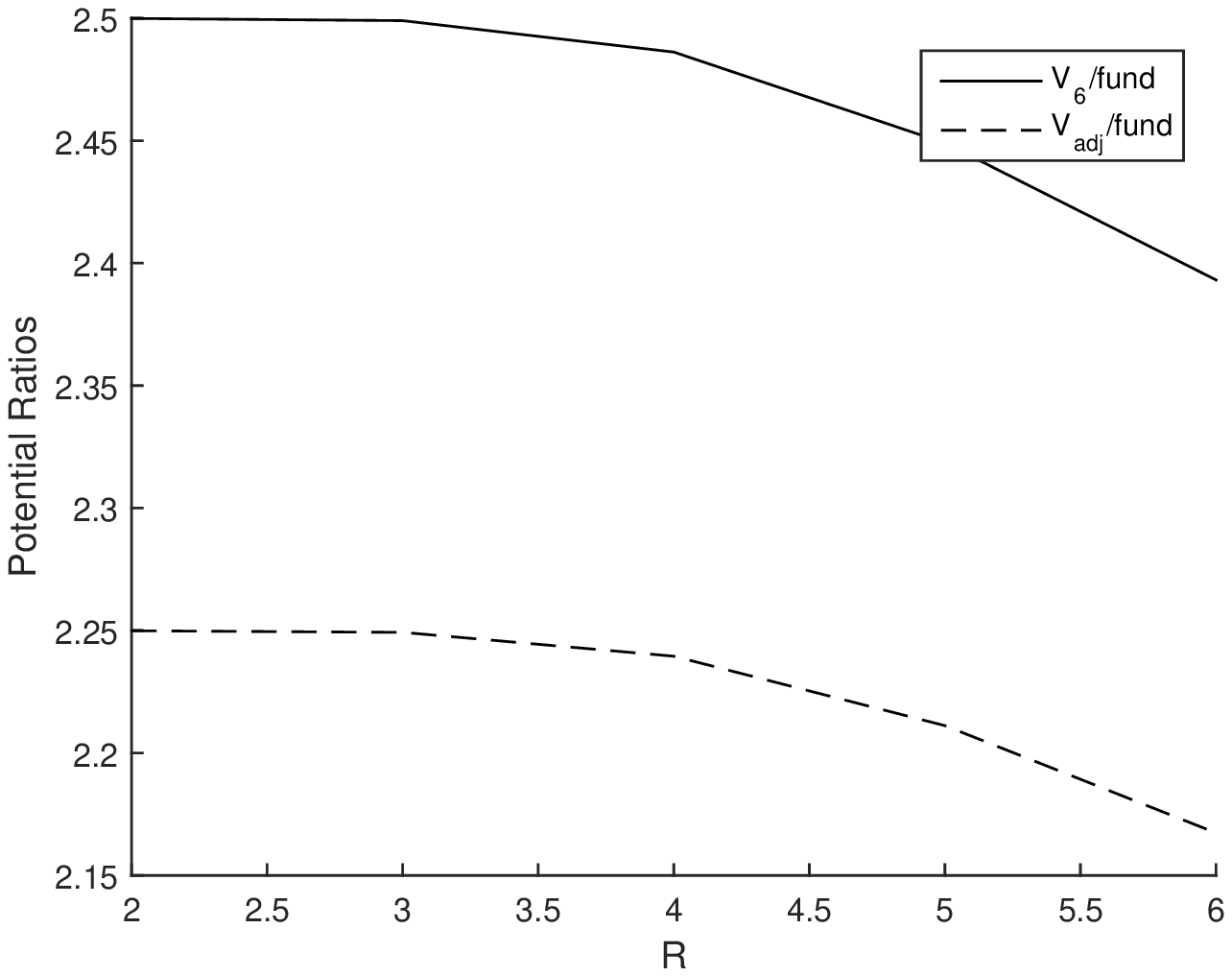}
\caption{Casimir ratios behaviour for the $SU(3)$ representations with non-symmetric vortex fluxes.}
\label{casimirtaksu3}
\end{center}
\end{figure*}

\begin{figure} [!ht]
\begin{center}
\includegraphics [scale=0.8] {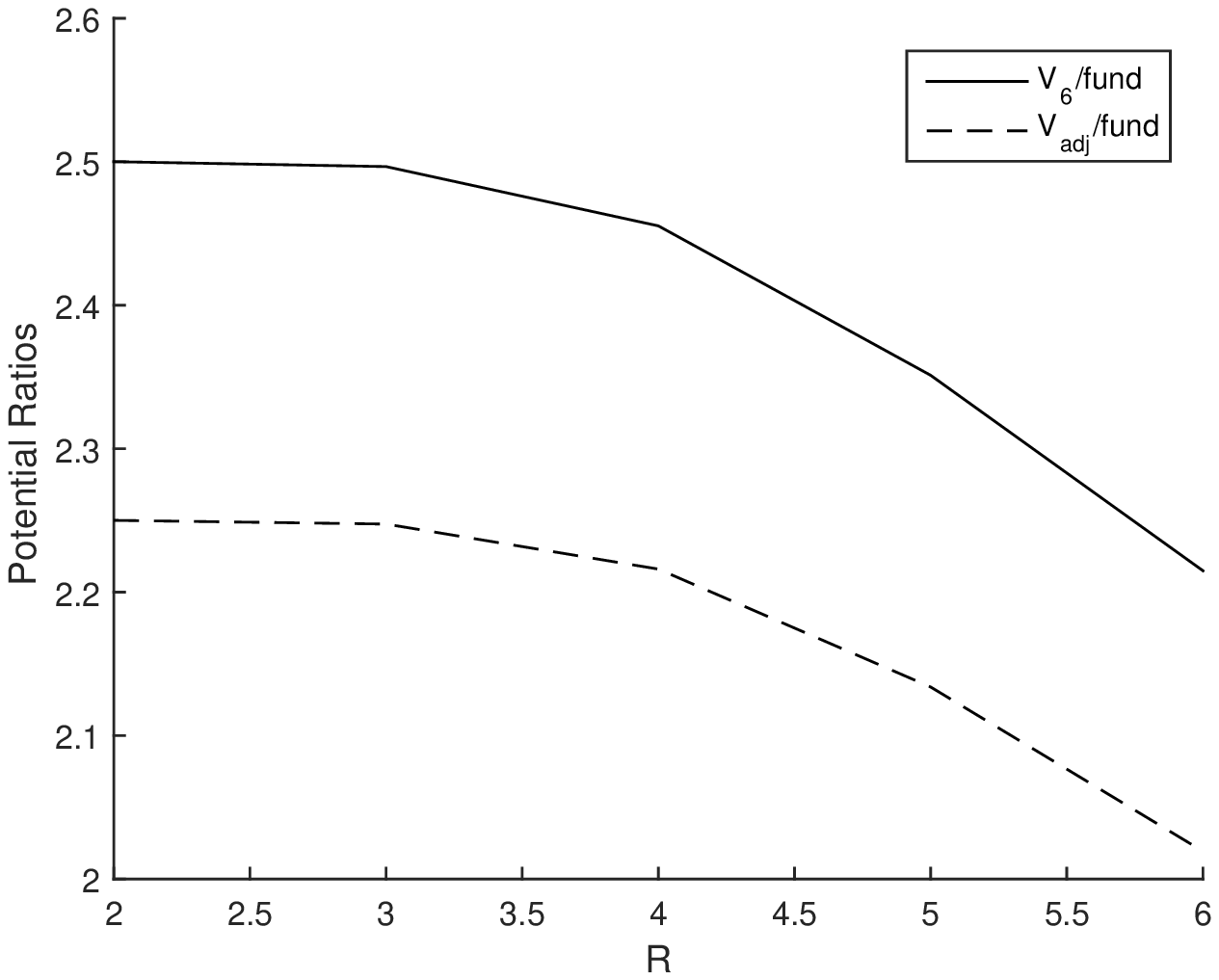}
\caption{Casimir ratios behaviour for the $SU(3)$ representations with arbitrary symmetric vortex fluxes.}
\label{casimirtwosu3}
\end{center}
\end{figure}

One of the characteristic properties of the quark potentials is Casimir scaling. According to this property the quark  linear potentials at the intermediate region should be proportional to the Casimir eigen value of the representation of the group. Consider $C_r$ as the Casimir eigen value of the representation $r$ and the quark potentials slope as $k_r$ then according to the Casimir rule: 

\begin{equation}
k_r=\frac{C_r}{C_F} k_F
\end{equation}
In which $F$ is the index for the fundamental representation. Figure \ref{taksu21} shows a close look at the potentials in this region for the representations of the $SU(2)$ with non-symmetric vortex fluxes. Figure \ref{twosu21} shows the potentials behaviour at intermediate region with two vortex fluxes. The Casimir ratios using the model can be obtained by dividing the potential values to the potential values of the fundamental representation for each distance at the intermediate region. Figure \ref{casimirtaksu2} shows the Casimir ratios for the $SU(2)$ by considering one vortex flux in the plane. Figure \ref{casimirtwosu2} shows Casimir ratios using two vortex fluxes with symmetric fluxes. The Casimir ratios are $\frac{C_3}{C_2}=\frac{8}{3} ,\frac{C_4}{C_2}=5$. Figure \ref{taksu31} shows the quark potentials for the $SU(3)$ representations at the intermediate region with non-symmetric vortex fluxes. Figure \ref{twosu31} shows the quark potentials for the $SU(3)$ representations at the intermediate region with symmetric vortex fluxes. Figure \ref{casimirtaksu3} shows the Casimir ratios for the $SU(3)$ group with one vortex flux.The Casimir ratios are $\frac{C_6}{C_3}=2.5 ,\frac{C_8}{C_3}=2.25$. Figure \ref{casimirtwosu3} shows Casimir ratios when two symmetric vortex are present in the plane. For these groups the effects of two symmetric vortex fluxes in the plane on the Casimir scaling is not very much and the Casimir scaling behaviour is present using these symmetric vortex fluxes. So, by considering symmetric vortex fluxes the concavity is removed with an acceptable Casimir scaling. 
  
  
\section{Conclusion}
Thick center vortex model and the domain structure picture of the QCD vacuum are the successful methods to explain the confinement problem. One of the important properties of the quark potentials is the downward concavity. As it is seen this property is violated using the thick center vortex model. Here it is shown that this property can be obtained using an arbitrary symmetric vortex flux. By considering such situation the concavity is removed from the quark potentials at the intermediate region and also an overall reduction of the concavity is observed for the quark potential. Applying the domain structure idea and considering a portion for the trivial center element in the quark potential behaviour leads to even better reduction of the concavity. The vortex thickness is related to the slope of the intermediate region of the quark potentials which are obtained from the thick center vortex model. The symmetric vortex flux leads to the elimination of the concavity within the intermediate region. By adding the trivial domain portion to the model and reduction of the concavity, it seems that the remaining concavity has other reason than the symmetric vortex profile. It may be related to the free parameters of the models such as $f,a,b$. Presence of two vortices in the plane can lead to interesting properties of the interaction between vortices. This can be studied in further research as the interaction between vortices.

\end{document}